\documentclass[aps,prb,twocolumn,superscriptaddress,nofootinbib]{revtex4-2}
\usepackage{miller}
\usepackage{xr}

\usepackage{graphicx}
\usepackage{amsmath}
\usepackage{xcolor}

\newcommand{\ie}{\textit{i.e.}}
\newcommand{\magnetite}{Fe$_3$O$_4$}
\newcommand{\hematite}{Fe$_2$O$_3$}
\footnotetext{These authors contributed equally to this work.}
\makeatletter
\newcommand*{\addFileDependency}[1]{
\typeout{(#1)}
\@addtofilelist{#1}
\IfFileExists{#1}{}{\typeout{No file #1.}}
}
\makeatother

\newcommand*{\myexternaldocument}[1]{%
\externaldocument{#1}%
\addFileDependency{#1.tex}%
\addFileDependency{#1.aux}%
}

\myexternaldocument{supplementary}

\usepackage{pdfpages} 
\usepackage{pgffor} 
\makeatletter
\AtBeginDocument{\let\LS@rot\@undefined}
\makeatother

\def\supplementfilename{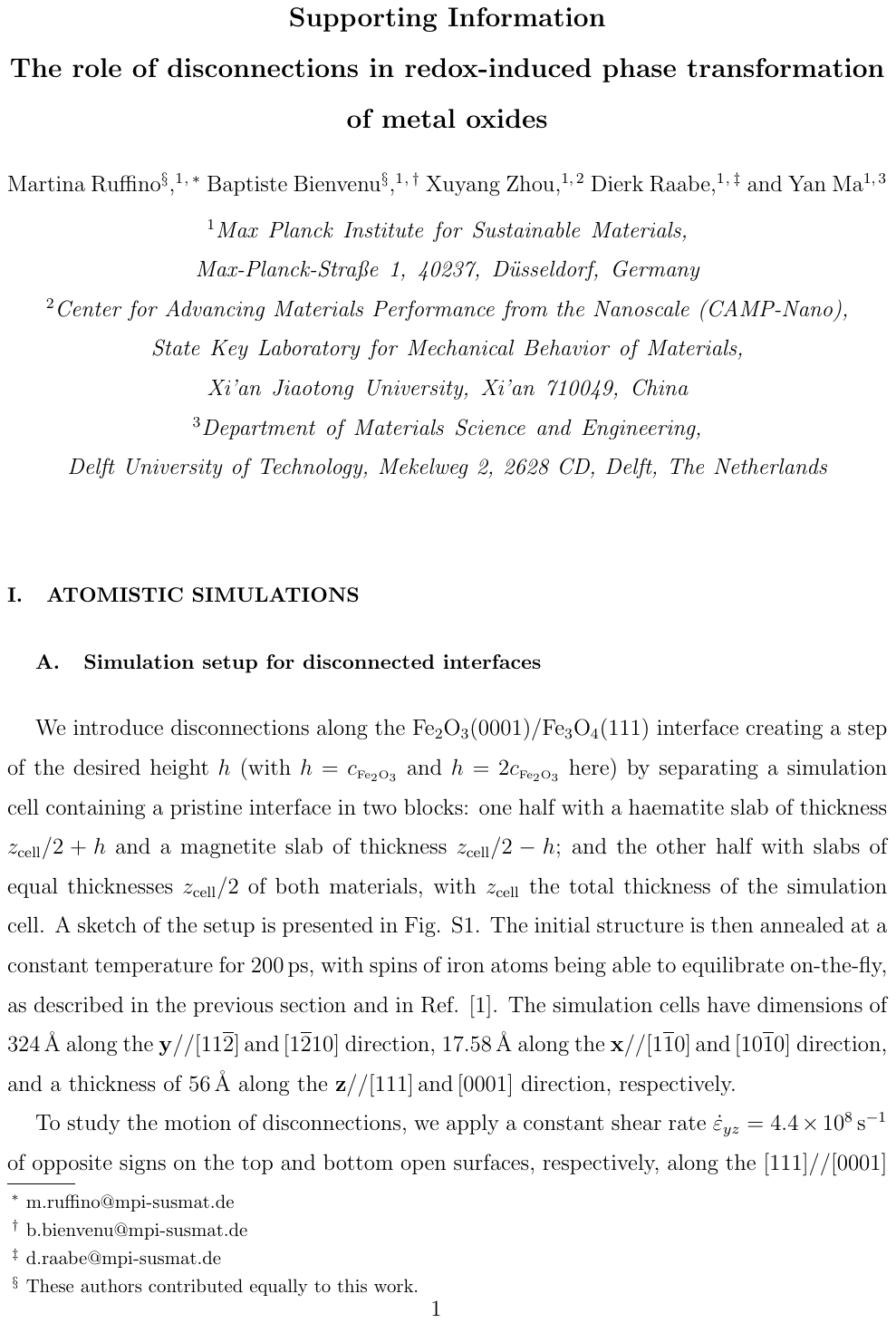}

\pdfximage{\supplementfilename}
\def\numbersupplementpages{\the\pdflastximagepages}

\newif\ifarXiv
\arXivtrue

\begin{document}

\title{The role of disconnections in redox-induced phase transformation of metal oxides}

\author{Martina Ruffino$^{\S}$}
\email{m.ruffino@mpi-susmat.de}
\affiliation{Max Planck Institute for Sustainable Materials, Max-Planck-Straße 1, 40237, Düsseldorf, Germany}
\author{Baptiste Bienvenu$^{\S}$}%
\email{b.bienvenu@mpi-susmat.de}
\affiliation{Max Planck Institute for Sustainable Materials, Max-Planck-Straße 1, 40237, Düsseldorf, Germany}
\author{Xuyang Zhou}
\affiliation{Max Planck Institute for Sustainable Materials, Max-Planck-Straße 1, 40237, Düsseldorf, Germany}
\affiliation{Center for Advancing Materials Performance from the Nanoscale (CAMP-Nano), State Key Laboratory for Mechanical Behavior of Materials, Xi’an Jiaotong University, Xi’an 710049, China}
\author{Dierk Raabe}
\email{d.raabe@mpi-susmat.de}
\affiliation{Max Planck Institute for Sustainable Materials, Max-Planck-Straße 1, 40237, Düsseldorf, Germany}
\author{Yan Ma}
\affiliation{Max Planck Institute for Sustainable Materials, Max-Planck-Straße 1, 40237, Düsseldorf, Germany}
\affiliation{Department of Materials Science and Engineering, Delft University of Technology, Mekelweg 2, 2628 CD, Delft, The Netherlands}

\date{16th September 2026}

\begin{abstract}
 Detailed atomic-level understanding of the phase transformation between magnetite ({\magnetite}) and haematite ({\hematite}) is lacking, despite widespread interest in redox reactions of metal oxide systems. While the magnetite-to-haematite transformation entails the addition of oxygen to the system and thus a net diffusive flux, haematite formations have been reported to grow along well-defined habits on the close-packed $\{0001\}_{\rm Fe_2 O_3}//\{111\}_{\rm Fe_3 O_4}$ planes. The propagation of the phase interface in the bulk is thus thought to maintain the oxygen sublattice fixed up to a shear, and to require long-range diffusion of iron ions. In this letter we propose interfacial steps with dislocation character, {\ie} disconnections, as the elementary defects propagating the transformation. We use the topological model of interfacial defects to predict three disconnection modes for the system, and we employ scanning transmission electron microscopy to ascertain their presence in a haematite/magnetite interface of a partially oxidised magnetite powder. Using atomistic simulations, we determine the equilibrium structures of the disconnections, and study their motion and how they accomplish the transformation, obtaining excellent agreement with atomic-resolution experimental observations.
\end{abstract}

\maketitle

Phase transformations in crystalline materials may be divided into diffusionless and diffusional \cite{christian1965}. Diffusionless transformations involve a change in the crystal structure without the presence of long-ranged diffusion, while diffusional transformations rely on diffusion across long distances, and may involve a change in chemical composition as well as crystal structure. Within this latter category fall redox-induced phase transformations, where the driving force for the transformation is provided by a gradient in the chemical potential induced by the removal (oxidation) or addition (reduction) of electrons; they are typically associated with the extensive rearrangement of atoms. Some, however, include the organised movement of atoms to a certain extent: this is the case of scale growth mediated by anion or cation diffusion on some metals, such as the epitaxial growth of NiO on Ni metal \cite{Sawhill1985}. A similar case is the oxidation of magnetite ($\rm{Fe}_3\rm{O}_4$, space symmetry group $Fd\bar{3}m$, inverse spinel structure) into haematite ($\rm{Fe}_2\rm{O}_3$, space symmetry group $R\bar{3}c$, corundum structure). During this transformation, nuclei of haematite grow into the magnetite matrix while keeping the oxygen sublattice largely fixed, up to a simple shear of the atoms to be achieved at the interface, such that long-range diffusion is confined to the iron ions \cite{Nie2013,Mccarty2014}. 

Theories on the partially organised nature of oxidation reactions in magnetite stemmed from the observation of lenticular formations of haematite penetrating into the magnetite bulk from the surface, resembling lenticular martensite \cite{Greig1935,Nie2013,Choisez2024}. The subject of magnetite oxidation has often been explored in association with that of reduction. Pre-oxidation of magnetite is known to improve its reducibility \cite{Wolfinger2021}, a topic of great interest in the context of reducing the carbon footprint of ironmaking \cite{Raabe2019}. Magnetite redox transformations are also of fundamental interest in topics such as corrosion prevention in steels \cite{Cornell2003}, hydrogen production from water-rock reactions \cite{Mayhew2013} and catalysis \cite{Ratnasamy2009}. While the microstructure of partially oxidised magnetite has been investigated, fundamental understanding of the atomic-scale mechanisms through which the transformation happens is still lacking. Such mechanisms are the focus of this article.

\begin{figure*}[!htb]
    \centering
    \includegraphics[width=0.9\textwidth]{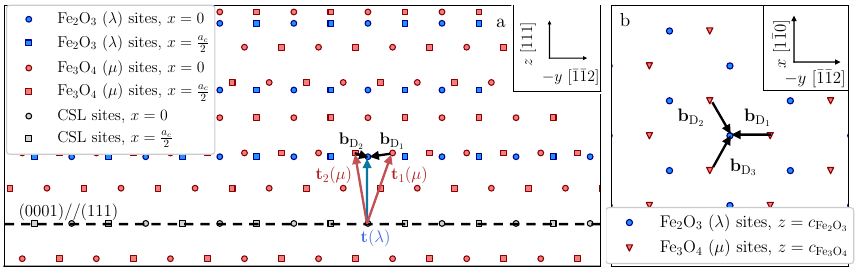}
    \caption{\label{fig:dichromatic}\textbf{a}: Coherent dichromatic pattern showing the interpenetrating oxygen sublattices of magnetite ($\mu$, red sites) and haematite ($\lambda$, blue sites), where the negative $x$-axis, $\hkl[-110]$, points out of the page; grey sites are overlapping $\lambda$ and $\mu$ sites, also known as coincident site lattice (CSL) points. The translation vectors $\mathbf{t}(\lambda)$, $\mathbf{t}_1(\mu)$ and $\mathbf{t}_2(\mu)$ that produce the disconnections as well as their Burgers vectors $\mathbf{b}_{\rm{D}_1}$ and $\mathbf{b}_{\rm{D}_2}$ are drawn. $a_c$ is the oxygen nearest neighbour spacing of the coherently strained pattern. \textbf{b}: Dichromatic pattern projected on $(0001)//(111)$, where the Burgers vector $\mathbf{b}_{\rm{D}_3}$ of the third disconnection is also visible. The $z$-axis is parallel to the positive $[111]$ direction, which points out of the page.}
\end{figure*}

Haematite has been reported to form in magnetite following the so-called Shoji-Nishiyama orientation relationship (S-N OR), where $\hkl(0001)_{\rm{Fe_2O_3}}//\hkl(111)_{\rm{Fe_3O_4}}$ and $\hkl[-1010]_{\rm{Fe_2O_3}}//\hkl[-110]_{\rm{Fe_3O_4}}$ (henceforth, the subscripts ``$\rm{Fe_2O_3}$" and ``$\rm{Fe_3O_4}$" will be dropped, as all 4-index and 3-index planes and directions are attributed to the haematite and magnetite crystals, respectively) \cite{Mccarty2014}. While haematite has rhombohedral symmetry, it is possible to consider its oxygen sublattice as a hexagonal close-packed (hcp) structure by neglecting the small shifts that occur in real crystals \cite{iida1957}, while the oxygen sublattice of magnetite has face-centred cubic (fcc) structure.
The structural oxygen transformation then amounts to a change in stacking sequence along $\hkl[0001]//\hkl[111]$ from fcc to hcp arrangement. Since the misfit between the $\hkl(0001)$ and $\hkl(111)$ oxygen lattice planes is small ($\sim 2\%$ \cite{Watanabe1995}), previous studies have proposed that the haematite/magnetite interface should be semi-coherent, with equally spaced interfacial misfit dislocations separating the coherent sections \cite{Hayes1981}.
Said misfit dislocations have also been identified as integral to boundary migration mechanisms in scaling reactions, particularly by Pieraggi, Rapp and coworkers \cite{Pieraggi1990,Loo1990}. For the opposite case of reduction of haematite to magnetite occurring on the same interface, Watanabe \textit{et al.} \cite{Watanabe1996} suggested that the transformation may take place through a slip-shear mechanism, where the change of stacking from hcp to fcc is achieved by the dissociation of haematite slip dislocations on the basal plane and their consecutive movement. Observations of the interface in projection along $[111]//[0001]$ have so far not shown evidence of this mechanism \cite{zhang2023}.

\begin{figure*}
    \centering
    \includegraphics[width=0.9\textwidth]{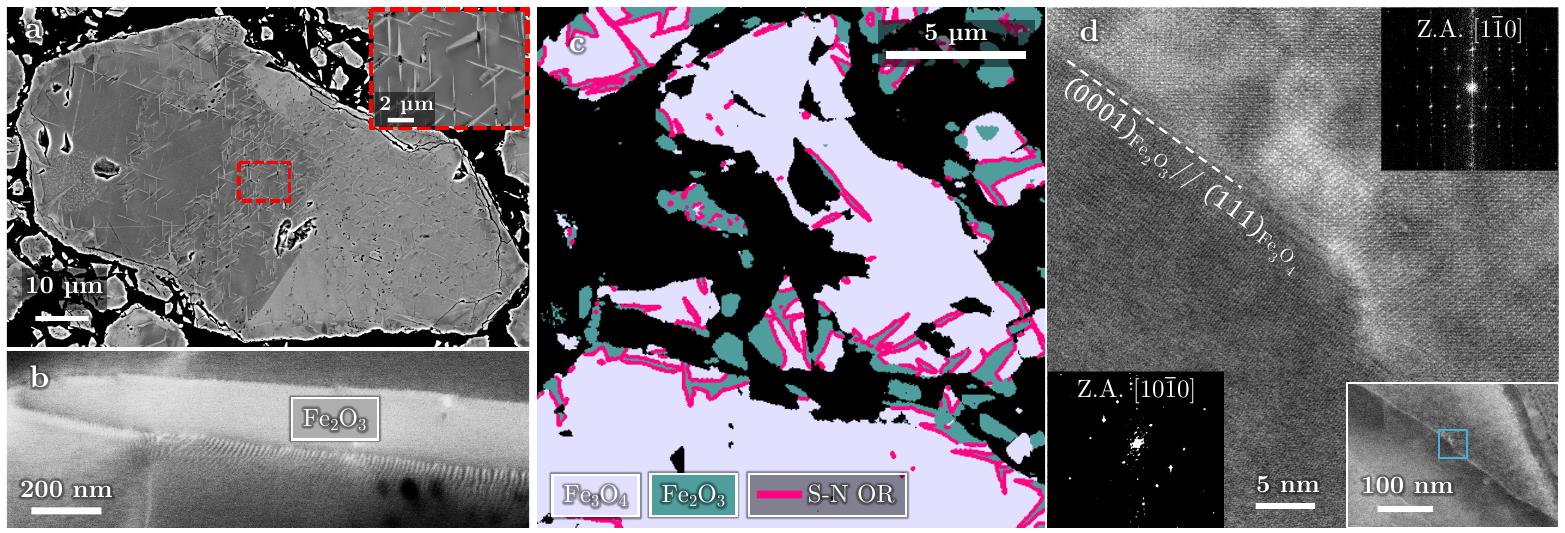}
    \caption{\label{fig:sem}Preliminary microstructure analysis of the partially oxidised magnetite powder. \textbf{a}: BSE image of a magnetite particle (dark gray contrast) containing haematite formations (light gray), magnified in the top-right inset. \textbf{b}: Electron channelling contrast image (ECCI) of a haematite plate. \textbf{c}: EBSD phase map with highlighted Shoji-Nishiyama interfaces (tolerance $\leq1^{\circ}$). \textbf{d}: Dark field scanning transmission electron microscopy (STEM) image of the faceted boundary of a lenticular haematite formation in magnetite; the top right and bottom left insets show fast Fourier transform images of the regions on either side of the interface, confirming that the sample was tilted to the $[10\bar{1}0]//[1\bar{1}0]$ zone axis. }
\end{figure*}

\section{Theory}
Atomic-scale understanding of the coupling of shear and diffusion for transformations in the \hematite-\magnetite\, system has so far not been reached. A type of interfacial defect combining these aspects is a disconnection, {\ie} a step with dislocation character \cite{Hirth1994}. Disconnection mechanisms have been proposed to adequately describe many diffusionless phase transformations \cite{Pond2003,Ma2007,Howe2009,Moritani2002,Ogawa2004}, diffusional transformations occurring on well-defined habit planes in such a way that the substitutional sites are conserved \cite{Howe2009,Howe1987,Pond2000}, as well as transformations involving a net diffusional flux \cite{Purdy2006}. Finally, disconnection motion was deemed dominant in scale growth on metals driven by anion diffusion \cite{Hirth1995,Howe2009}. Here we will show how, while haematite growth in magnetite occurs through cation diffusion, the transformation can still be accomplished by disconnections.

In order to predict available disconnection modes for our system we employ the topological model \cite{Pond1988}. The creation of a bicrystal, {\ie} the object produced by bringing together two crystals at an interface, is a symmetry-breaking process, because not all the symmetry operations of the two crystals survive in the bicrystal. These lost symmetry operations characterise admissible interfacial defects. 
For a defect with dislocation content, the mathematical operation that describes it, {\ie} the Burgers vector, is obtained by combining translation operations belonging to the lattices of the two crystals that were lost upon creation of the bicrystal.
Labelling the crystals $\mu$ and $\lambda$, and the transformation relating them $\mathbf{P}$, the Burgers vector of the dislocation in the coordinate system of $\lambda$ is given by $\mathbf{b} = \mathbf{t}(\lambda)-\mathbf{P}\mathbf{t}(\mu)$, where $\mathbf{t}(\lambda)$ and $\mathbf{t}(\mu)$ are translation vectors belonging to each crystal lattice.
Disconnections are further characterised by the height $h$ of the associated step, which is obtained as the overlap of the heights of the step in the $\mu$ and $\lambda$ crystals \cite{Hirth1996}, {\ie} the smaller between $h_{\lambda}=\hat{\mathbf{n}}\cdot\mathbf{t}(\lambda)$ and $h_{\mu}=\hat{\mathbf{n}}\cdot\mathbf{t}(\mu)$, where $\hat{\mathbf{n}}$ is the unit normal of the interface plane.
The Burgers vector and step height of a disconnection can be visualised by constructing a coherent dichromatic pattern for the system, {\ie} the interpenetrating lattices belonging to the $\mu$ and $\lambda$ crystals.

This is shown for the haematite/magnetite system in Fig. \ref{fig:dichromatic}\,a, where the lattice sites represent the oxygen sublattices of the two crystals.
As the observed interface plane of the transformation is $\hkl(0001)//\hkl(111)$, the two lattices were coherently strained, such that the plane of grey sites (overlapping $\mu$ and $\lambda$ sites) was obtained; here, the common $\hkl[-1010]//\hkl[-110]$ direction points out of the page.
The Burgers vectors of all admissible dislocations and disconnections are then given by the vectors connecting $\mu$ and $\lambda$ sites.
It is seen that the smallest possible step height of a disconnection is $h=c_{\scriptscriptstyle \rm{Fe_2O_3}}$, where $c_{\scriptscriptstyle \rm{Fe_2O_3}}$ is the length of the $c$-axis of the haematite oxygen sublattice, while two choices of Burgers vector are possible, labelled $\mathbf{b}_{\rm{D_1}}$ and $\mathbf{b}_{\rm{D_2}}$. A third admissible Burgers vector may be visualised in projection onto the $\hkl(0001)//\hkl(111)$ plane, labelled $\mathbf{b}_{\rm{D_3}}$ (see Fig. \ref{fig:dichromatic}\,b). It is seen that the disconnection with $\mathbf{b}_{\rm D_1}$ has pure edge character, while the disconnections with $\mathbf{b}_{\rm D_{2,3}}$ have mixed edge and screw character.
In the orthonormal coordinate system where $\hat{\mathbf{x}}=\frac{[1\bar{1}0]}{|[1\bar{1}0]|}$, $\hat{\mathbf{y}}=\frac{[{1}{1}\bar{2}]}{|[{1}{1}\bar{2}]|}$ and $\hat{\mathbf{z}}=\frac{[111]}{|[111]|}$, the Burgers vectors of the disconnections are given by
\begin{align}
    \mathbf{b}_{\rm{D_1}} &= \big[0,\frac{a_c}{\sqrt{3}},\sqrt{\frac{8}{3}}a_{\scriptscriptstyle \rm{Fe_3O_4}}-c_{\scriptscriptstyle \rm{Fe_2O_3}}\big],\\
    \mathbf{b}_{\rm{D_2}} &= \big[-\frac{a_c}{2},-\frac{\sqrt{3}a_c}{6},\sqrt{\frac{8}{3}}a_{\scriptscriptstyle \rm{Fe_3O_4}}-c_{\scriptscriptstyle \rm{Fe_2O_3}}\big],\\
    \mathbf{b}_{\rm{D_3}} &= \big[\frac{a_c}{2},-\frac{\sqrt{3}a_c}{6},\sqrt{\frac{8}{3}}a_{\scriptscriptstyle \rm{Fe_3O_4}}-c_{\scriptscriptstyle \rm{Fe_2O_3}}\big],
\end{align}
with $a_c = \frac{a_{\scriptscriptstyle \rm{Fe_3O_4}}+a_{\scriptscriptstyle \rm{Fe_2O_3}}}{2}$ the oxygen nearest neighbour spacing of the coherently strained bicrystal (in determining the sign of the Burgers vectors, the SF/RH convention is followed, where the line sense of the disconnection points into the page). Motion of any of these disconnections on the $\hkl(0001)//\hkl(111)$ plane transforms the fcc oxygen sublattice of magnetite into the hcp sublattice of haematite, as their Burgers vector accomplishes the necessary macroscopic shear, while the remaining shuffles needed to restore the lattice are controlled by the step.
We further propose that rearrangement and local diffusion of iron ions also takes place through the step riser, such that a single defect may encompass the atomic level mechanism responsible for the transformation.

\begin{figure*}[!htb]
    \hspace{0mm}
    \includegraphics[trim = 1mm 1mm 1mm 1mm, clip, width=1\linewidth]{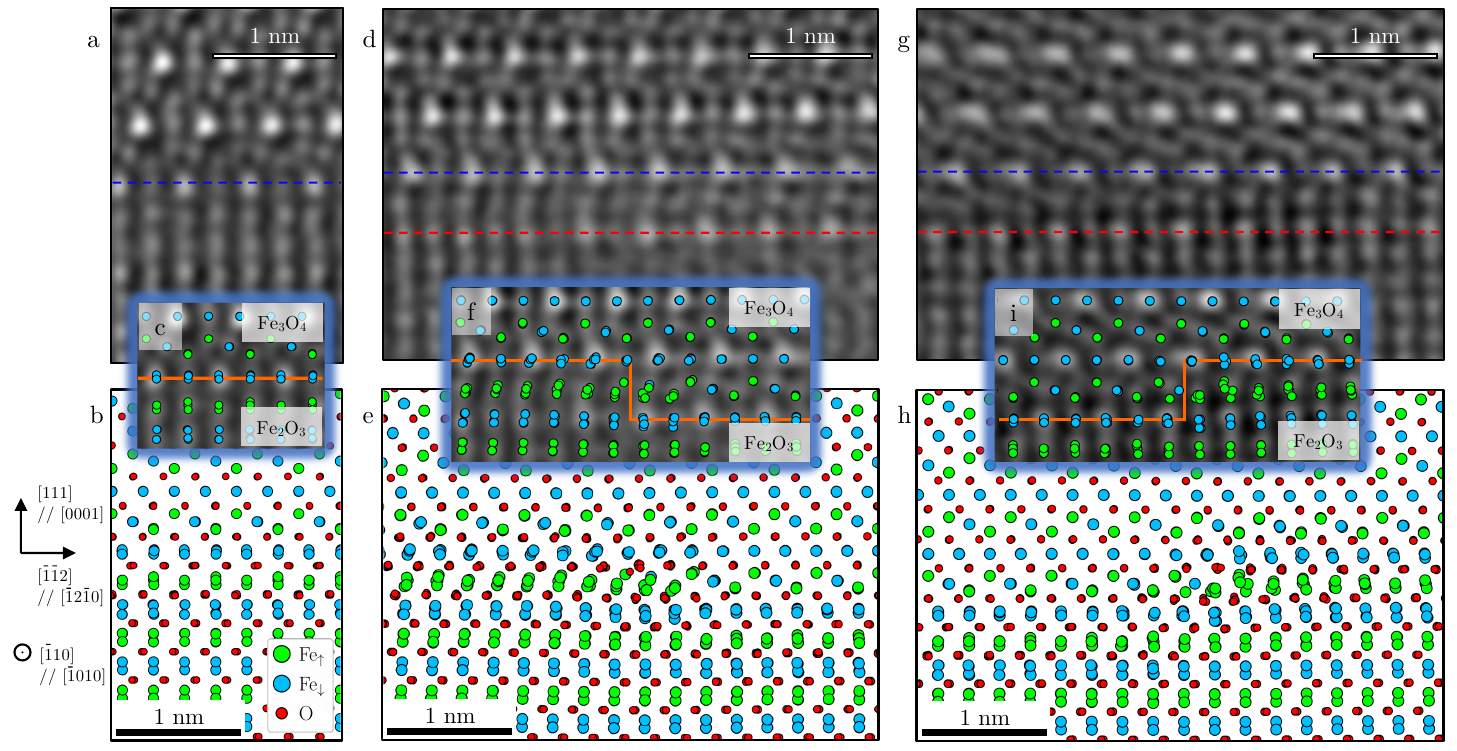}
    \caption{Comparison between experimental STEM images and equilibrium structures predicted by atomistic simulations for a: \textbf{a}-\textbf{c} pristine portion of the {\hematite}\hkl(0001)/{\magnetite}\hkl(111) interface; \textbf{d}-\textbf{f}: portion containing a $\mathbf{b}_{\rm{D_1}}$ disconnection; \textbf{g}-\textbf{i}: portion containing a $\mathbf{b}_{\rm{D_{2/3}}}$ disconnection.
    In each presented STEM image (\textbf{a}, \textbf{d} and \textbf{g}), the position of the interface is indicated by horizontal dashed lines, the lower one in red and the upper one in blue. Structures predicted by atomistic simulations (\textbf{b}, \textbf{e} and \textbf{h}) show Fe ions in green and blue, depending on the sign of their spins, and O ions in red. In the insets (\textbf{c}, \textbf{f} and \textbf{i}), showing superpositions of STEM images and atomistic simulations, the interface is located by an orange line. In both STEM images and atomistic simulations, the same exact area is covered.
    }
    \label{fig:fig3_TEM_atomistic}
\end{figure*}

\section{Results}
We validate the proposed theory using both experiment and atomistic simulations. Here we investigate the $\text{\magnetite}\rightarrow\text{\hematite}$ transformation in bulk samples, as opposed to studies of redox reactions in nano-sized samples present in the literature \cite{zhang2023}. The starting material is a magnetite powder oxidised in air at $700^{\circ}$C for 30 min (powder composition as that of ``powder B" in ref. \cite{Zheng2022}; further details on the material and experimental techniques can be found in Section IV. Materials and Methods). A magnetite particle is shown in backscattered electron (BSE) imaging mode in Fig. \ref{fig:sem}\,a, where lenticular haematite formations can be distinguished by lighter contrast.
An electron backscatter diffraction (EBSD) phase map (Fig. \ref{fig:sem}\,c) reveals that haematite and magnetite orientations follow the S-N OR as reported by previous studies (with a tolerance of $\leq1^{\circ}$). Finally, electron channelling contrast imaging (ECCI) of a haematite plate within the magnetite matrix (Fig. \ref{fig:sem}\,b) shows strain contrast at the interface that may be attributed to dislocations, as well as the non-planar nature of the interface. Scanning transmission electron microscopy (STEM) investigations were conducted on a lamella prepared via focused ion beam and tilted to the common $\hkl[10-10]//\hkl[1-10]$ zone axis. A dark field image of a large facet in the haematite/magnetite interface is shown in Fig. \ref{fig:sem}\,d. The facet separates terraces of planar interface parallel to the $\hkl(0001)//\hkl(111)$ plane, revealing that the non-planar boundary seen via ECCI in Fig. \ref{fig:sem}\,b is in fact stepped.

We further probe the interface by descending to the atomic scale. We first focus on the pristine (\ie~ defect-free) interface, of which a high-angle annular dark field (HAADF) STEM image is presented in Fig. \ref{fig:fig3_TEM_atomistic}\,a.
Only the iron ions are clearly visible in the HAADF images, from which the position of the oxygen ions can be inferred. To determine the exact structure of the interface, we use atomistic simulations with a recently developed Atomic Cluster Expansion (ACE) potential for the Fe-O system \cite{Bienvenu2025}. The equilibrium structure of the lowest-energy pristine interface is presented in Fig. \ref{fig:fig3_TEM_atomistic}\,b (more details are presented in Supporting Information, Fig. S2).
The bridging sketch presented in Fig. \ref{fig:fig3_TEM_atomistic}\,c shows the very good agreement found between the atomic positions in the HAADF image and those predicted by atomistic simulations. 

HAADF imaging is similarly used to identify disconnections in the interface. Two such defects are shown in Fig. \ref{fig:fig3_TEM_atomistic}\,d and g. In order to characterise them, closed Burgers vector circuits are drawn around the defects and mapped to the dichromatic pattern, taken as the reference state (shown in Supplementary Figure S5).
The Burgers vector content is thus found as the closure failure of the circuit in the reference state, and corresponds to $-\mathbf{b}_{\rm{D_1}}$ for the disconnection in Fig. \ref{fig:fig3_TEM_atomistic}\,d, and either $\mathbf{b}_{\rm{D_2}}$ or $\mathbf{b}_{\rm{D_3}}$ for that in Fig. \ref{fig:fig3_TEM_atomistic}\,g. As predicted by the theory, the step height of both disconnections is $h=c_{\scriptscriptstyle \rm{Fe_2O_3}}$.

\begin{figure}[!htb]
    \centering
    \hspace{0mm}
    \includegraphics[trim = 1mm 1mm 1mm 1mm, clip, width=0.8\linewidth]{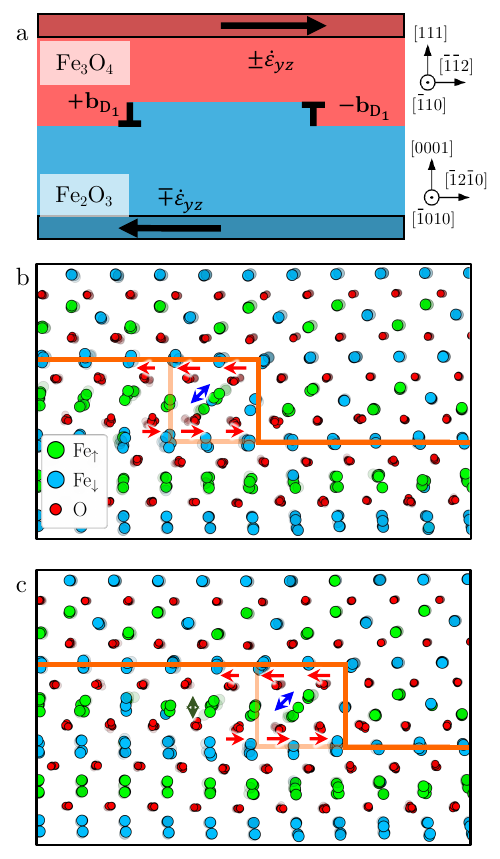}
    \put(-185,230){\colorbox{white}{\normalsize $\varepsilon_{1}-\varepsilon_{0}$}}
    \put(-185,105){\colorbox{white}{\normalsize $\varepsilon_{2}-\varepsilon_{1}$}}
    \caption{Snapshots of a molecular dynamics trajectory showing the elementary motion of a $\mathbf{b}_{\rm{D_1}}$ disconnection under an applied shear strain at 1000\,K. \textbf{a}: Sketch of the setup, showing the strain rate $\dot{\varepsilon}_{yz}$ applied to the top and bottom regions of the simulation cell. \textbf{b,c} Differential images between two snapshots corresponding to $\varepsilon_{1}=4.6\,\%$ and $\varepsilon_{0}=3.9\,\%$ (\textbf{b}), $\varepsilon_{2}=5.1\,\%$ and $\varepsilon_{1}=4.6\,\%$ (\textbf{c}). On each differential image, shaded (full) colour circles show the atoms in their previous (updated) positions. The position of the interface before (after) motion of the disconnection is represented by an orange shaded (full) line.}
    \label{fig:fig4_motion}
\end{figure}

We then use atomistic simulations to characterise the core structure of these defects, by constructing simulation cells containing each one of the two observed disconnections, $\mathbf{b}_{\rm{D_1}}$ and $\mathbf{b}_{\rm{D_{2/3}}}$, which are then annealed at 600\,K and subsequently relaxed. We present on Fig. \ref{fig:fig3_TEM_atomistic}\,e and h the predicted relaxed core structure of both disconnections, with bridging sketches presented on Fig. \ref{fig:fig3_TEM_atomistic}\,f and i showing the agreement found between atomistic simulations and HAADF images.
Apart from the vicinity of the disconnection cores, where both iron and oxygen sublattices are slightly deformed, the two sides of the interface between the haematite and magnetite layers retain the same structure as observed and predicted for pristine regions, up to shifts of the iron layer along the $[10\bar{1}0]//[1\bar{1}0]$ direction, which do not change the interfacial energy .

We now analyse the motion of the observed disconnections, focusing on the pure edge $\mathbf{b}_{\rm{D_1}}$ Burgers vector (Fig. \ref{fig:fig3_TEM_atomistic}\,d-f). To activate the motion of the disconnection, we shear the lattice by applying a constant strain rate $\dot{\varepsilon}_{yz}$, as sketched in Fig. \ref{fig:fig4_motion}\,a, along a molecular dynamics trajectory at 1000\,K. The simulation cell contains a disconnection dipole, where the $+\mathbf{b}_{\rm{D_1}}$ and $-\mathbf{b}_{\rm{D_1}}$ disconnections move in opposite directions under the applied strain. 
Selected snapshots showing the elementary steps for the motion of the $-\mathbf{b}_{\rm{D_1}}$ disconnection are presented in Fig. \ref{fig:fig4_motion}\,b,c after quenching down to 0\,K for better visualization. We make sure this relaxation step does not disrupt the dynamics as observed along the trajectory by smoothly relaxing atomic positions using the FIRE algorithm \cite{Bitzek2006} until all Cartesian components of atomic forces fall below $0.1$\,eV/{\AA}.

During the simulation, the motion of the $-\mathbf{b}_{\rm{D_1}}$ disconnection causes the oxygen sublattice to be deformed by shearing of the layer corresponding to the step height (see the orange lines in Fig. \ref{fig:fig4_motion}\,b,c), resulting in the desired change of the oxygen stacking sequence from fcc to hcp (see red arrows). Simultaneously, iron ions located at the core of the disconnection rearrange themselves from mixed tetrahedral-octahedral rows, resembling magnetite, to doubled rows of same spin orientation, resembling haematite (see blue arrow). Once each sublattice has deformed and rearranged around its core, the disconnection has moved one elementary step. As the disconnection moves another step further (see Fig. \ref{fig:fig4_motion}\,c), the deformation of the iron sublattice caused by the core of the disconnection partially resorbs (see green arrow), leaving behind a pristine interface, and a completely reordered lattice resembling haematite where the disconnection core was located before.

Additionally, we performed similar simulations with a $\mathbf{b}_{\rm{D_{2/3}}}$ disconnection dipole, during which we also observed their motion upon application of a strain. We note that the critical shear strain at which the disconnections started moving is lower for the pure edge $\mathbf{b}_{\rm{D_1}}$ disconnection than for the mixed $\mathbf{b}_{\rm{D_{2/3}}}$. As the total Burgers vector content and elastic strain energy is minimised when the disconnections travel in a train or sequence $(\mathbf{b}_{\rm D_1},\mathbf{b}_{\rm D_2},\mathbf{b}_{\rm D_3})$, we also simulated a disconnection pile-up with total Burgers vector $-(\mathbf{b}_{\rm D_1}+\mathbf{b}_{\rm D_2})$, and analysed its motion under the applied strain, obtaining good agreement with HAADF observation of the same structure; details are presented in Supporting Information (Fig. S6), where it is shown that the pile-up spontaneously decomposes into elementary disconnections with $h=c_{\scriptscriptstyle\rm Fe_2 O_3}$.

\section{Discussion}

Overall, two driving forces are expected to contribute to the $\text{\magnetite}\rightarrow\text{\hematite}$\, phase transformation as mediated by disconnection motion. A mechanical force is necessary to initiate motion of the dislocation part of the disconnection, which leads to local iron ion rearrangement through the step riser. Such a driving force is expected to be provided by the presence of coherency strains in the interface; indeed, local interfacial stresses have been shown to be high in phase-field simulations of phase transformations in iron oxide systems (of the orders of tens of GPas if no plastic relaxation occurs \cite{BAI2022}). The chemical potential driving force is then necessary to prompt long-range diffusion of iron ions away from the transformation front, thus restoring proper haematite stoichiometry. In our simulations, since the chemical potentials of both atomic species are fixed, the change in concentration associated with the formal transformation from magnetite to haematite cannot be fully described. For a disconnection such as that presented in Fig. \ref{fig:fig4_motion}, moving to the right, the necessary change in concentration would be $-1$ Fe ion per unit of volume (where one volume unit is defined as $3\sqrt{3}a_c^2c_{\scriptscriptstyle\rm Fe_2 O_3}$, as detailed in Supporting Information).
In order to assess the ease of removal of these ions, we calculated Fe vacancy formation energies in a layer of non-stoichiometric haematite produced by the transit of the disconnection in Fig. \ref{fig:fig4_motion}, and compared them with those of Fe ions in the cell containing only the pristine interface (Fig. S4). 
In the undeformed configuration, the lowest formation energies are those of Fe ions in the pristine interface, the removal of which would be insufficient to support the transformation. Conversely, in the transformed cell negligible formation energies are found for Fe vacancies belonging to the three Fe layers comprised by the transformed oxygen sublattice. This indicates that once the lattice has been deformed by a moving disconnection, the necessary change in stoichiometry is more likely to occur than in the pristine interface. The flux of iron ions away from the transformation front is then to be supported by an iron ion vacancy source, expected to be in the magnetite. This is due to the superior electrical conductivity of magnetite as compared with haematite ($\sim10^{2}$ \cite{Cornell2003} and $\sim 10^{-11}$ \cite{engel2014} S/cm at room temperature, respectively), such that the flow of an opposite electron current associated with the necessary iron ion diffusion is greatly facilitated. Thus, we expect the mechanical and chemical driving forces of the transformation to be highly coupled.

In summary, we have presented a disconnection-based mechanism for the $\text{\magnetite}\rightarrow\text{\hematite}$ phase transformation. We used the topological model to identify three interfacial disconnections, and verified their existence via HAADF-STEM investigation of a partially oxidised magnetite sample. We employed atomistic simulations to characterise the haematite/magnetite interface and to determine the core structure of the disconnections, obtaining excellent agreement with our high-resolution HAADF observations. Finally, we used molecular dynamics to shear the simulation cell and trigger the motion of the disconnections, observing them propagate the phase transformation, and discussed the associated Fe ion diffusion necessary to complete the transformation. We thus established disconnection motion as an effective mechanism for haematite growth in magnetite, providing novel insights into redox-induced phase transformations. \\

\section{Materials and Methods}
\subsection*{Experimental methods}For the present study, we used \magnetite~  powder with $\ge 96\%$ purity and $90\%$ of particles measuring less than 63 $\rm \mu$m in size (``powder B" in ref \cite{Zheng2022}). 1 g of the powder was oxidised in an open-ended horizontal tube furnace at 700$^{\circ}$C, allowing for air exchange, for 30 min (final oxidation degree $76.5\pm3.7\%$). The partially oxidised samples were mounted in a conductive resin and ground with SiC abrasive papers, followed by polishing using diamond microparticle and silica nanoparticle suspensions. Microstructure investigations were conducted using using a Zeiss Merlin high-resolution scanning electron microscope in backscattered electrons (BSE) and electron channelling contrast imaging (ECCI) \cite{ZAEFFERER2014} modes, operated at 10 kV and 2.2 nA.  Electron backscatter diffraction (EBSD) maps were collected using an EDAX DigiView 5 camera and EDAX APEX software. The Kikuchi pattern data sets were analysed by spherical indexing using EDAX OIM Matrix v.9 software \cite{LENTHE2019}. A lamella (3 $\rm{\mu m}\times5~\rm{\mu m}\times2~\rm{\mu m}$) was prepared from a \magnetite~ particle via in-plane lift out in the dual-beam FEI Helios NanoLab 600 focused-ion-beam (FIB); it was then mounted on a copper grid and thinned up to a thickness of $\sim 100$ nm. High-angle annular dark field scanning (HAADF) images were captured using a Cs probe-corrected FEI Titan Themis 60-300 microscope operating at 300 kV. All HAADF images were processed via Fourier filtering and Gaussian smoothing.

\subsection*{Atomistic simulations}

Atomistic simulations presented in this work have been performed using the LAMMPS code \cite{Thompson2022}, with a recently developed Atomic Cluster Expansion (ACE) potential for the Fe-O system \cite{Bienvenu2025} to model interactions between atoms. As detailed in the original publication supporting the development of the ACE potential, the model has an explicit account of magnetic degrees of freedom of the iron atoms by defining different magnetic species. All molecular dynamics (MD) simulations are performed with on-the-fly equilibration of the magnetic order of the iron sublattice by performing semi-grand canonical Monte Carlo swap attempts between the two Fe$_{\uparrow}$ and Fe$_{\downarrow}$ species along the trajectory. Further details concerning the simulation cells are provided in Supporting information.

\bibliography{bibliography}


\appendix

\makeatletter
\@addtoreset{equation}{section}
\@addtoreset{figure}{section}
\@addtoreset{table}{section}
\makeatother

\clearpage

\ifarXiv
    \foreach \x in {1,...,\numbersupplementpages}
    {
    	\clearpage
        \includepdf[pages={\x}, fitpaper=true]{\supplementfilename}
    }
\fi

\end{document}